\documentclass[prc,reprint]{revtex4-1}
\usepackage{graphicx}   
\usepackage{latexsym}   

\begin{document}
\title{Directed flow, a signal for the phase transition in Relativistic Nuclear Collisions?}

\author{J. Steinheimer, J. Auvinen, H. Petersen, M. Bleicher and H. St\"ocker}

\affiliation{Frankfurt Institute for Advanced Studies, Ruth-Moufang-Str. 1, 60438 Frankfurt am Main, Germany}
\affiliation{Institut f\"ur Theoretische Physik, Goethe Universit\"at Frankfurt, Max-von-Laue-Strasse 1, D-60438 Frankfurt am Main, Germany}
\affiliation{GSI Helmholtzzentrum f\"ur Schwerionenforschung GmbH, Planckstr.~1, D-64291 Darmstadt, Germany}

\begin{abstract}
The sign change of the slope of the directed flow of baryons has been predicted as a signal for a first order phase transition within fluid dynamical calculations. Recently, the directed flow of identified particles has been measured by the STAR collaboration in the beam energy scan (BES) program. In this article, we examine the collision energy dependence of directed flow $v_1$ in fluid dynamical model descriptions of heavy ion collisions for $\sqrt{s_{NN}}=3-20$ GeV. The first step is to reproduce the existing predictions within pure fluid dynamical calculations. As a second step we investigate the influence of the order of the phase transition on the anisotropic flow within a state-of-the-art hybrid approach that describes other global observables reasonably well. We find that, in the hybrid approach, there seems to be no sensitivity of the directed flow on the equation of state and in particular on the existence of a first order phase transition. In addition, we explore more subtle sensitivities like e.g. the Cooper-Frye transition criterion and discuss how momentum conservation and the definition of the event plane affects the results. At this point, none of our calculations matches qualitatively the behavior of the STAR data, the values of the slopes are always larger than in the data. 

\end{abstract}

\maketitle
\section{Introduction}

The anisotropic flow of particles has been an interesting observable, since data from the first heavy ion collisions became available at the Bevalac. The deflection of the produced particles in the reaction plane (defined as the plane between impact parameter and beam direction) can be quantified by the so called directed flow, $v_1$. At very low beam energies of $E_{\rm{lab}}<1$ GeV per nucleon, the rotation of the system will lead to a strong overall directed flow coefficient, that has been observed and understood within fluid dynamical calculations \cite{Baumgardt:1975qv,Hofmann:1976dy,Stoecker:1980vf}.

At very high beam energies, as they are achieved at the Large Hadron Collider (LHC) and the Relativistic Heavy Ion Collider (RHIC), the slope of the traditional directed flow is close to zero at midrapidity due to the almost perfect transparency of the colliding nuclei. The small negative slope of charged particles (mostly pions) at top RHIC energy can be explained within a fluid dynamical model and a slightly tilted initial state \cite{Bozek:2010bi} as well as a hadronic transport model \cite{Snellings:1999bt}. In the last 3 years more studies where focused on odd flow coefficients related to initial state fluctuations. The so called rapidity-even $v_1$/directed flow was defined to quantify the dipole moment generated by fluctuations in the initial transverse density profile.
In the present study, we are solely interested in the traditional rapidity-odd directed flow, that forms independent of initial fluctuations.

At intermediate colliding energies, studied at the beam energy scan program at RHIC, the future Facility for Antiproton and Ion Research (FAIR) and the former AGS-SPS experiments, the systematic study of directed flow is thought to be more interesting. Within fluid dynamical calculations, it has been predicted that the slope of the directed flow of baryons will turn negative and then positive again as a function of energy if a first order phase transition is present. This means that more protons (most of the baryons at lower beam energies are protons) are emitted in direction opposite to the spectators than aligned with them. This effect, called "antiflow" or "collapse of flow", has been attributed to a softening of the equation of state (EoS), during the expansion, due to a first order phase transition \cite{Rischke:1995pe,Csernai:1999nf,Brachmann:1999xt,Stoecker:2004qu}, leading to a rotation or tilt of the fireball in the reaction plane \cite{csernai14}.
The corresponding measurements of the NA49 collaboration \cite{Alt:2003ab} for the directed flow of protons had insufficient statistics to draw definite conclusions. Recently, the STAR collaboration has measured the predicted qualitative behavior of the slope of the net-proton directed flow as a function of beam energy which turns negative and then positive again \cite{Adamczyk:2014ipa}. Since the early predictions were made with exclusively fluid dynamical models, which over predicted all other flow components, the goal of our study is to understand the EoS dependence of directed flow within more modern transport approaches.

First, we validate the qualitative predictions within a pure fluid dynamical calculation and confirm that with a first order phase transition the proton $v_1$ slope has the expected qualitative behavior, including a dip below zero. As in the previous studies, this sign change happens at much lower beam energies than what STAR has measured. In Section \ref{section_purehydro} we explore the influence of the freeze-out criterion on this result (isochronous compared to iso-energy density) and show the relation to the time evolution of the directed flow. Then we perform the calculation within the Ultrarelativistic Quantum Molecular Dynamics (UrQMD) hybrid approach with a more realistic treatment of the initial state and final stages employing non-equilibrium hadron-string transport. In this calculation the sensitivity of the directed flow to the equation of state is less obvious. Finally, in Section \ref{section_discussion} we point out additional issues that need to be addressed, before a clear conclusion can be drawn.    

\section{The Equations of State}

In the following we will study the effect of the equation of state of hot and dense nuclear matter on the directed flow measures in relativistic nuclear collisions. In particular we want to know whether the slope of the directed flow, as function of rapidity, is sensitive to the order of the QCD phase transition. We therefore have to compare two different scenarios. One where the QCD transition is of first order and one where it is a crossover.

For the first order transition scenario we will employ a well known Maxwell construction which has been used in several investigations on the effect of the EoS \cite{Rischke:1995mt}. The Maxwell construction is used to connect a mean field type SU(2)$_f$ hadronic model (HM) and a Bag Model EoS (BM) that consists of free quarks and gluons. The conditions for a Maxwell construction are the equality of the thermodynamic variables temperature $T_{BM}=T_{HM}$, baryochemical potential $\mu_{BM}=\mu_{HM}$ and pressure $p_{BM}=p_{HM}$. As a result of the construction one obtains a single phase system inside the coexistence region of the transition.
For simplicity, in the following, we will refer to the constructed EoS only as the Bag Model EoS (BM).
 
Due to the Maxwell construction the iso-thermal speed of sound $c_s^{IT}$ essentially vanishes,
and also the isentropic speed of sound $c_s^{IE}$ drops considerable inside inside the coexistence region. Note that the Maxwell construction only accounts for the so called ''softening'' of the equation of state, due to the phase transition, and it lacks important features associated with a first order phase transition, e.g. a region of mechanical instability or the surface tension \cite{Randrup:2009gp,Steinheimer:2012gc}. However since we are only interested in the effect of the softening on the bulk dynamics the Maxwell constructed EoS will suffice for our current investigations.

Alternatively we will use an equation of state which follows from the combination of a chiral hadronic model with a constituent quark model \cite{Steinheimer:2011ea}, later referred to as the $\chi$-over equation of state. This EoS gives a reasonable description of lattice QCD results at vanishing net baryon density, including a smooth crossover from a confined hadronic phase to a deconfined quark phase. This crossover continues into the finite density region of the phase diagram for essentially all densities relevant for the present investigations. 

We therefore are able to compare the fluid dynamics of systems which always evolve through a first order transition to those which always evolve through a smooth crossover.  

\section{Cold Nuclear Matter Initialization}
\label{section_purehydro}

Early studies on the directed flow in relativistic nuclear collisions suggested the 'collapse of flow' to be a possible signal for a first order phase transition in dense nuclear matter. In particular one extracted the net x-momentum per nucleon $p_x^{dir}/N$, defined in the direction of the impact parameter, in a given rapidity window from fluid dynamical simulations as: 
\begin{equation}\label{pxdir}
	p_x^{dir}/N=\frac{\int{\rho_B(r) m_N v_x(r) dr}}{\int{\rho_B(r)}dr}
\end{equation}
where $\rho_B(r)$ and $v_x(r)$ are the local net baryon density and fluid velocity and $m_N$ is the nucleon mass. It was found that, when the bag model equation state, with a strong first order phase transition, was used, $\left.d p_x^{dir}/d y\right|_{y=0}$, i.e. the slope of the directed net momentum with rapidity, would be negative for collisions where the system remains in the mixed phase for a considerable time \cite{Rischke:1995pe}. 

As a first step we want to reproduce these results with the above described bag model EoS, that includes a Maxwell construction from a hadronic to a quark gluon plasma phase. Furthermore we also use the  $\chi$-over EoS that only shows a crossover and is consistent with lattice data at $\mu_B=0$, to see if the observed 'anti-flow' is unique for a first order transition, i.e. a very strong softening.

We use the 1fluid SHASTA algorithm, which is an ideal 3+1d fluid dynamic code described in \cite{Rischke:1995ir} for all calculations. 

\begin{figure}[t]	
\includegraphics[width=0.5\textwidth]{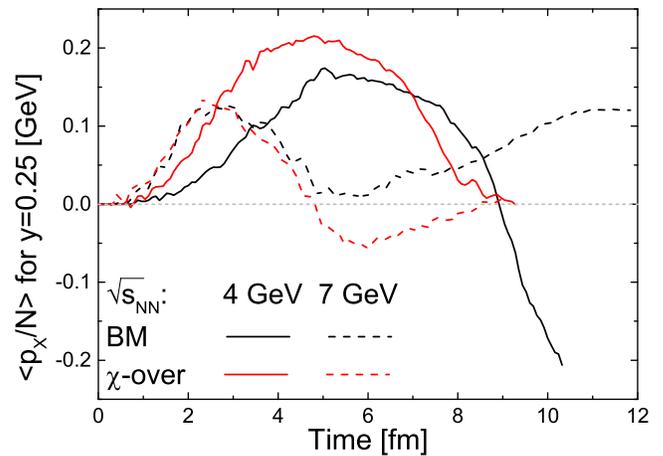}	
\caption{[Color online] The time dependence of $\left\langle p_x^{dir}/N \right\rangle$ at fixed rapidity window $y=0.25 \pm0.05$ from the ideal 1-fluid calculations with a bag model and crossover EoS. We show two different beam energies, $\sqrt{s_{NN}}=4$ and $7$ GeV, solid and dashed lines respectively. There is an evident non-monotonic time dependence of $\left\langle p_x^{dir}/N \right\rangle$.
}\label{f1}
\end{figure}		

In early investigations \cite{Rischke:1995pe} the full collision was simulated within ideal fluid dynamics. As a consequence the two colliding nuclei have to be described as two homogeneous density distributions colliding head on. In this so called ''cold nuclear matter initialization'' no distinct nucleons exist, but two distributions of cold, locally equilibrated, nuclear matter. We therefore initialize two energy- and baryon density distributions, according to boosted Woods-Saxon profiles with a central density of saturated nuclear matter $\rho_0 \approx 0.16 \rm{fm}^{-3}$, corresponding to the two Au nuclei with a given center of mass energy, at impact parameter $b=8$ fm. The simulation is started at a point in time just before the two nuclei first make contact. In the early stage of the collision the kinetic energy of the nuclei is then rapidly stopped and transformed into large local densities.
From the consecutive fluid dynamical simulation we can extract $\left\langle p_x^{dir}/N \right\rangle$, according to eq. (\ref{pxdir}) as a function of time and at fixed rapidity, in the center of mass frame. Figure \ref{f1} shows a comparison of the time dependent directed momentum per nucleon as extracted from the pure 1-fluid simulation at two different beam energies. A noticeable non-monotonic dependence of the directed flow on time can be observed, due to the angular momentum of the fireball, and we expect the final result to depend considerably on the decoupling time of the evolution. 

\begin{figure}[t]	
\includegraphics[width=0.5\textwidth]{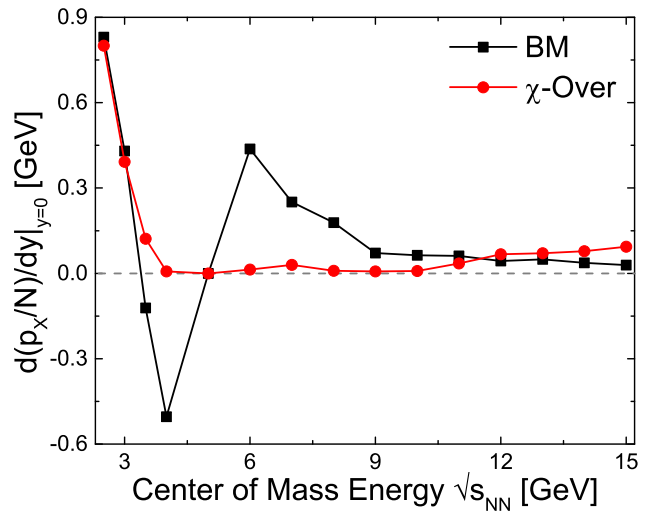}	
\caption{[Color online] Beam energy dependence of the directed flow slope around mid rapidity. Extracted from the ideal 1-fluid calculations with a bag model (black) and $\chi$-over EoS (red) for Au+Au collisions, with an impact parameter of $b=8$ fm. 
}\label{f2}
\end{figure}

A typical transition point, from the fluid dynamical phase to the final hadronic decoupling, used in most recent simulations \cite{Petersen:2008dd,Petersen:2010cw} is roughly four times the nuclear ground state energy density $\epsilon_0$. 
The slope of $p_x^{dir}/N$ is obtained by a linear fit, to the rapidity dependent $p_x^{dir}/N(y)$ between $-0.5<y<0.5$, at the time when all cells of the calculation are below that criterion. In figure \ref{f2} we compare the beam energy dependence of the slopes from both possible equations of state, the first order transition and crossover scenarios.

As can be seen, we reproduce the predicted negative slope of the directed flow around  $\sqrt{s_{NN}}= 4$ GeV when a first order transition is present \cite{Rischke:1995pe}. The crossover EoS also shows a minimum over a range of $\sqrt{s_{NN}}= 4 \ - \ 10$ GeV, however the slope always remains positive.

Already in the early studies it was noted that the quantity extracted with equation (\ref{pxdir}) is not directly comparable to experimental measured, identified particle $v_1$, defined as:
\begin{equation}\label{v1}
	v_1=\left\langle cos(\phi- \Psi_{RP}) \right\rangle
\end{equation}
where $\Psi_{RP}$ is the reaction plane angle and $\phi$ the transverse angle of a particular particle. The average is usually performed over all particles in all events, in a given rapidity bin. In order to transform the fluid dynamical fields into real particles we will use the Cooper Frye prescription \cite{Cooper:1974mv} on a pre defined hypersurface. The hypersurface, extracted from the unique fluid dynamical final state, is then used to sample a large number of hadronic final states which are independently evolved in the UrQMD transport model. 
Because the slope of the directed x-momentum was extracted from the fluid dynamical simulation at a fixed time we will first use a isochronous hypersurface for our particle production. 

\begin{figure}[t]	
\includegraphics[width=0.5\textwidth]{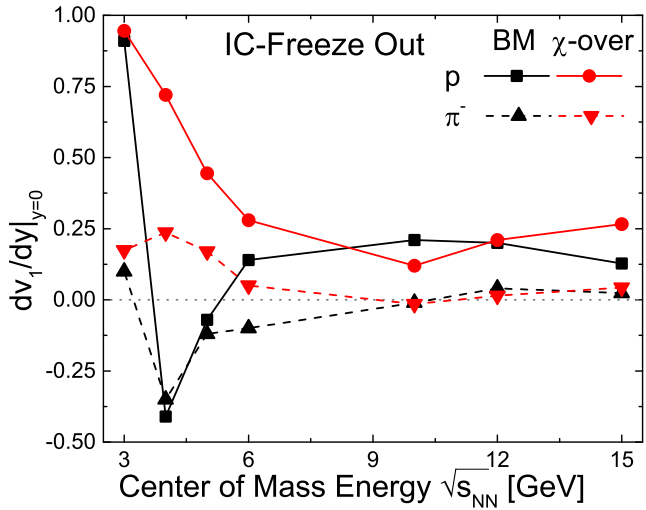}	
\caption{[Color online] Beam energy dependence of the $v_1$ slope of protons and negatively charged pions around mid rapidity extracted from the ideal 1-fluid calculations with a bag model (black) and $\chi$-over EoS (red). For particle production we applied a Cooper-Frye prescription on a iso-chronous hypersurface.
}\label{f3}
\end{figure}

The resulting slopes of the directed flow around mid rapidity (fitted for $-0.5<y<0.5$), for different particle species, are shown in figure \ref{f3}. 
Again, the negative slope is observed in the first order transition scenario around $\sqrt{s_{NN}}= 4$ GeV for protons and pions. The calculation with the $\chi$-over EoS shows only a broad minimum in the $d v_1/d y$ slope for protons and pions. However it always remains positive. The softening of the two EoS therefore leads to a minimum of the directed flow slope, but not always to a negative ''anti-flow''. Also the position of the minimum in beam energy varies with the EoS, as the crossover leads to a softening also at larger densities, resulting in a very shallow minimum at larger beam energy. 
   
In figure \ref{f4} we show the same quantity as in figure \ref{f3}, but this time we use an iso-energy density hypersurface for the transition in the subsequent hadronic afterburner. To construct this hypersurface we employ the Cornelius hypersurface finder \cite{Huovinen:2012is}, which has been already successfully used in previous studies \cite{Auvinen:2013ira}. The minimum in the extracted $d v_1/dy$ slopes occurs again at the same beam energies as with the isochronous freeze out, however the proton $v_1$ slope remains now always positive, even when we use the EoS with the large softening due to the phase transition.
As shown in figure \ref{f1} the directed flow, at a given rapidity, shows a non monotonic time dependence. The positive proton $v_1$ slope in the iso-energy density freeze out scenario can therefore be regarded as the result of an effective shortening of the fluid dynamical 'low viscosity' phase as compared to the iso-chronous freeze out.
 
 It is noteworthy that in both discussed cases the slope of the proton $v_1$ always turns positive again once the beam energy is increased above the 'softest point' of the EoS and that the pion directed flow shows always the same qualitative behavior as the proton flow.

\section{Hybrid Model} 
\label{section_hybrid}
Until now we have assumed that the colliding systems are in local equilibrium from the beginning of the collision, in order to apply ideal fluid dynamics also for the initial interpenetration phase. This had the advantage that we could use different equations of state also for the initial phase, which leads to different compression dynamics and subsequently has an impact on the directed flow. However, the assumption of local equilibrium is certainly not justified for the very early stage of a nucleus-nucleus collision. In this stage a non-equilibrium approach is better suited to describe the early time dynamics. One example of such an approach is studied in this section. 

The UrQMD hybrid approach, described in detail in \cite{Petersen:2008dd}, was introduced to combine the advantages of a Boltzmann transport approach with fluid dynamics. Because the UrQMD model is used for the initial interpenetration stage of the collision, the effective equation of state during that stage is defined by the microscopic properties of the model, i.e. a purely hadronic phase. As stated in Section \ref{section_purehydro}, the fluid dynamical evolution is realized using the SHASTA algorithm, while the initial state before equilibrium and the final state with hadronic rescatterings and decays are computed using UrQMD. 

\begin{figure}[t]	
\includegraphics[width=0.5\textwidth]{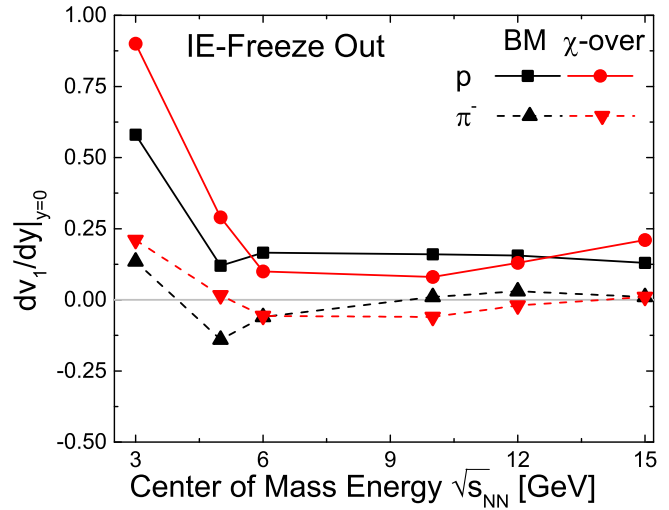}	
\caption{[Color online] Slope of $v_1$ of protons and pions around mid rapidity extracted from the ideal 1-fluid calculations with a bag model (black) and $\chi$-over EoS (red). For particle production we applied a Cooper-Frye prescription on a iso-energy density hypersurface.
}\label{f4}
\end{figure}		

In the hybrid simulations, the transition from initial transport to fluid dynamics happens when the two colliding nuclei have passed through each other: $t_{\mbox{start}}=\frac{2R}{\sqrt{\gamma_{CM}^2-1}}$, where $R$ represents the nuclear radius and $\gamma_{CM}$ 
is the Lorentz factor in the center-of-mass frame of the colliding nuclei. The transition from fluid dynamics back to transport happens on an iso-energy density $\epsilon = 4\epsilon_0 \approx 0.6 \ \rm{GeV}/\rm{fm}^3$ surface, which is constructed using the Cornelius hypersurface finder.

\begin{figure}[t]	
\includegraphics[width=0.5\textwidth]{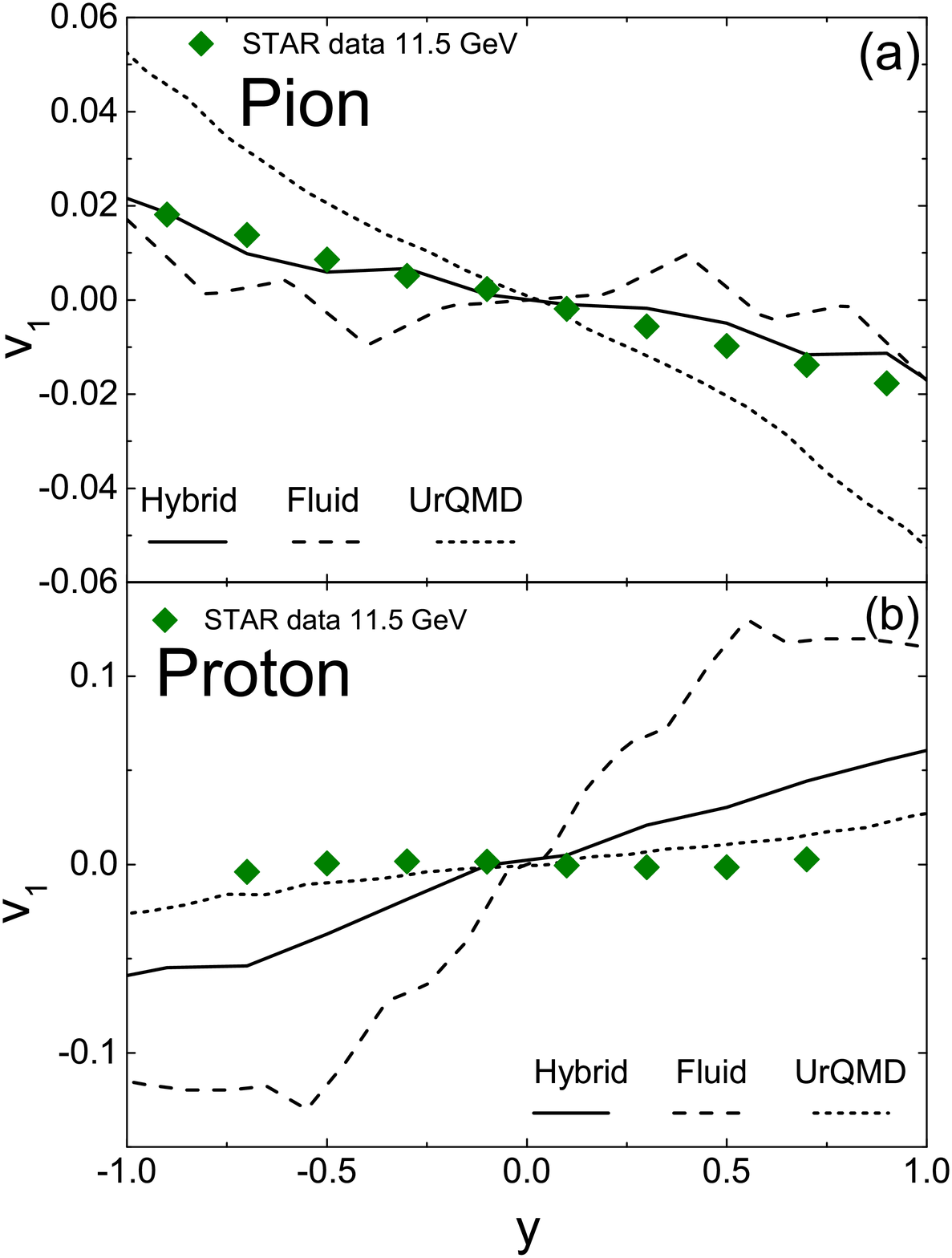}	
\caption{[Color online] Comparison of pion (a) and proton (b) $v_1(y)$  from the various models, for a beam energy of $\sqrt{s_{NN}}= 11.5$ GeV, compared with data \cite{Adamczyk:2014ipa}. Here we always used the $\chi$-over EoS in the fluid dynamical phase.
}\label{f5}
\end{figure}

The directed flow was calculated using events with impact parameter $b=4.6-9.4$ fm, to approximate the $(10-40)\%$ centrality range of the STAR data. As seen in Figure \ref{f5} (b), the hybrid model overestimates the directed flow as function of rapidity for protons at $\sqrt{s_{NN}}=11.5$ GeV, in comparison to the experimental data and the standard UrQMD result. However, for charged pion $v_1$ at the same collision energy the hybrid model results agree with experimental data better than standard UrQMD or the pure fluid dynamical simulation (Fig. \ref{f5} (a)). All the hybrid model calculations, up to $\sqrt{s_{NN}}\approx 16$ GeV, reproduce the qualitative feature observed at lower SPS energy, that the proton $v_1$ has the opposite sign of the pion $v_1$.

\begin{figure}[t]	
\includegraphics[width=0.5\textwidth]{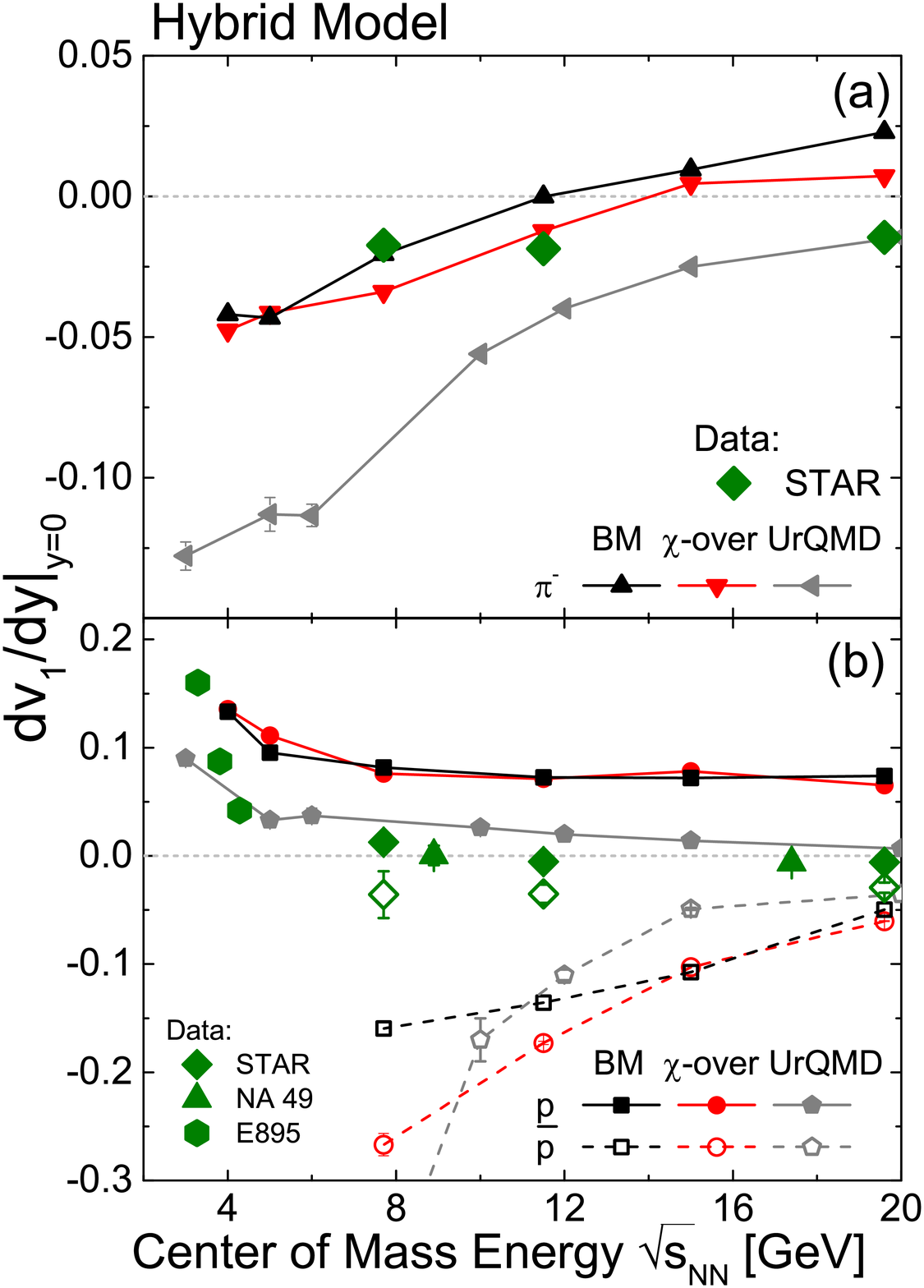}	
\caption{[Color online] Slope of $v_1$ of negatively charged pions (a) and protons and anti-protons (b) around mid rapidity extracted from the hybrid model calculations with a bag model and crossover EoS. We compare with standard UrQMD and experimental data \cite{Adamczyk:2014ipa,Alt:2003ab,Liu:2000am}.
}\label{f7}
\end{figure}

%
The full beam energy dependence for the hybrid model results of the midrapidity $v_1$ slope for negatively charged pions and protons/antiprotons is shown in figure \ref{f7} (a) and (b) respectively. Both proton and antiproton slopes are overestimated for the whole examined collision energy range, while $\left.dv_1/dy\right|_{y=0}$ for pions agrees with the data at lower collision energies but changes sign at $\sqrt{s_{NN}} \approx 10$ - $15$ GeV, which is not supported by the STAR data. While the difference between the investigated equations of state was already rather small in the pure fluid results (Fig. \ref{f4}), the two EoS are completely indistinguishable in the hybrid simulations.

For comparison we also present the standard UrQMD results as grey lines. The qualitative behavior is very similar to the hybrid model results. The standard UrQMD appears to better describe the experimental proton data, however.

\begin{figure}[t]	
\includegraphics[width=0.5\textwidth]{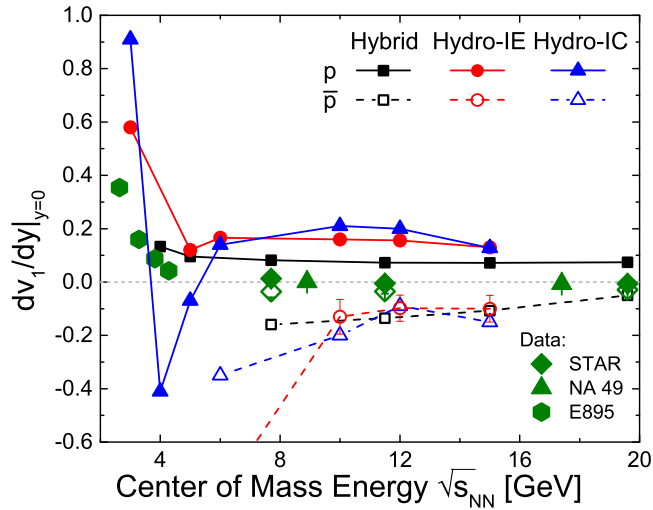}	
\caption{[Color online] Comparison of results for pure hydro and hybrid model calculations with a first order EoS. Shown is the slope of $v_1$ of protons and anti protons around mid rapidity compared to experimental data.
}\label{f9}
\end{figure}

Figure \ref{f9} shows the comparison of the hybrid and the pure hydro model results for the midrapidity $dv_1/dy$ for protons and antiprotons, where the bag model equation of state is used. Both approaches give significantly too large slopes compared to the STAR data. As noted already in Section \ref{section_purehydro}, the proton $dv_1/dy$ changes sign only for the model with isochronous Cooper-Frye hypersurface.
  
No model calculation seems to capture the qualitative experimental trend showing the directed flow slope for all particles turning negative at some point and approaching 0 from below. Also the overall magnitude seems to be strongly overestimated.

\section{Discussion}
\label{section_discussion}

As observed in previous studies \cite{Auvinen:2013sba,Huovinen:2012is,Petersen:2006vm}, the hybrid model (and to some extend also standard UrQMD) usually shows a reasonable agreement with experimental particle spectra, as well as the second and third order flow coefficients, at the energies investigated in this paper. Furthermore it has been shown \cite{Csernai:2011gg,Retinskaya:2012ky,Gardim:2011qn,Teaney:2010vd} that fluid dynamical simulations can quite successfully account for the rapidity even $v_1$ moment, which is caused by initial state fluctuations.

The strong deviation of the directed flow of all models, as compared to data, noted in the previous section is therefore surprising and requires a further discussion about the possible sources of differences in the $v_1$ extracted from the model as compared to the experiment.

One particular difference for example lies in the determination of the reaction plane angles $\Phi_{RP}$ used in equation (\ref{v1}). In our study the reaction plane (RP) angle is always defined to be zero along the impact parameter axis (x-axis). Experiments determine a $v_1$ event plane (EP) using certain assumptions. Usually the EP for the directed flow is defined along the vector of the projectile and target spectator transverse momentum motion (defined also by measurement) \cite{Agakishiev:2011id,Abelev:2013cva}. In an ideal scenario a model study would also define the EP in such a way. 

However, this is not possible here, as the spectators in the hybrid model, by definition, do not obtain a momentum correlated with the RP and in the pure hydro calculation are $100 \%$ correlated with the true reaction plane. In the hybrid model, as in standard UrQMD, the particles usually defined as spectators obtain only a random total momentum, caused by the finite net Fermi momentum of the spectators. As a result both spectators have essentially uncorrelated transverse momenta and are not correlated to the RP. Furthermore this spectator momentum should be balanced by momentum transfered to the fireball, in the initial state. Due to our definition of $v_1$ with respect to the true RP this "conserved" momentum does not contribute to the extracted $v_1$ as it would if we used a EP defined as in experiment. 

Finally a more realistic scenario should not only consider the momentum transfer from the spectator to the fireball (and vice versa) from the random Fermi momenta but also from likely correlations and binding of the cold nuclei \cite{Alvioli:2010yk}. We know experiment measures a finite $p_x$ for the spectators, but not it's origin. This momentum must be balanced, so the fireball should have a momentum contributing opposite to the "naive" $v_1$ by definition, which is lacking in our model study.

Future quantitative investigations on the correlations between the spectators as well as their average transverse momentum might help to constrain model uncertainties arising from the incomplete description of spectator-fireball momentum transfer. 

\section{Conclusions}
We have presented model simulation results on the directed flow of identified particles in nuclear collisions of beam energies ranging from $\sqrt{s_{NN}}=3$-$20$ GeV. To describe the strongly interacting systems created in these collisions we used different approaches, combining hadronic transport and ideal fluid dynamics. 

We find that the pure fluid dynamical approach can reproduce older findings \cite{Rischke:1995pe,Brachmann:1999xt} which predicted a negative slope of the proton directed flow if a strong first order phase transition is present in the EoS. However we also find that this 'anti-flow' is only observed if , and only if the full dynamics, expansion and initial compression, is treated fully in the ideal fluid dynamics model. In the idealized 1-fluid case the slope of the directed flow becomes positive at beam energies above the ''softest point'' of the EoS, just as observed in the 3-fluid simulations \cite{Brachmann:1999xt} and hybrid model.

When we apply a more realistic freeze out procedure and a hadronic transport model for the initial state we observe a positive slope of the proton directed flow for all beam energies under investigation.
Comparing our results to experimental data we find that essentially all models, including the standard hadronic transport UrQMD, cannot even describe the qualitative behavior, observed by experiment, of the proton directed flow. All models severely overestimate the data, even though other observables, like the radial or elliptic flow, are usually well described within these models.

Therefore the measured negative slope of the directed flow is not explained. The calculated directed flow is very sensitive to details in the description of the initial state, the freeze out prescription (and f.o. time) as well as the method of determining the event plane. 
The definition of the event plane in experiment, as well as the momentum transfer between the spectators and the fireball is not properly treated in the present model calculations.
These issues need to be addressed, before definite conclusions about the relation between the slope of directed flow and the EoS (including a phase transition) can be made.

\section{Acknowledgments}
We would like to thank Declan Keane, Mike Lisa, Paul Sorensen, Huan Huang and Nu Xu for interesting discussions about the STAR data.
This work was supported by GSI and the Hessian initiative for excellence (LOEWE) through the Helmholtz International Center for FAIR (HIC for FAIR). H. P. and J. A. acknowledge funding by the Helmholtz Association and GSI through the Helmholtz Young Investigator grant VH-NG-822. The computational resources were provided by the LOEWE Frankfurt Center for Scientific Computing (LOEWE-CSC).



			\end{document}